# Development of X-ray spectrometer automatic adjustment system based on global optimization algorithm

Fei Zhan Email：zhanf@sari.ac.cn
https://orcid.org/0000-0003-2757-4816
Shanghai Advanced Research Institute, Chinese Academy of Sciencesan

## Abstract

In high energy resolution X-ray spectroscopy beamlines of synchrotron radiation facilities, it is important to keep X-ray spectrometer operating in optimal
conditions. The adjusting process is normally very time consuming due to the irregular light source beam point, and it is difficult to get global optimum. This study aims to develop an intelligent adjusting system based on global optimization algorithm for high energy resolution X-ray spectroscopy beamline commissioning. First of all, based on the two dimensional experimental data, automatic adjustment process was established. Then the automatic optimization was applied to adjust X-ray spectrometer practically, and upgraded iteratively.
Online tests results show that this automatic adjustment process converges to the optimal solution quickly, and the convergence time is about several hundred steps, more efficient than manual optimization process. After automatic adjustment, we can get correct X-ray spectrum based on the adjusted spectrometer.

**Key words**：
global optimization algorithm, X-ray spectrometer,automatic adjustment

X-ray spectrometer based on spherical bent crystal has become one of the important experimental devices based on synchrotron radiation source. Its importance has been widely recognized by researchers in various fields. An important application field of X-ray spectrometer is to develop high energy resolution X-ray spectroscopy experimental method on synchrotron radiation sources, such as high energy resolution fluorescence detected X-ray absorption spectroscopy (HERFD-XAS), X-ray emission spectroscopy (XES), resonant inelastic X-ray scattering (RIXS) and X-Ray Raman Scattering Spectroscopy (XRS) etc. Now many X-ray spectrometer had been built on synchrotron radiation beamlines (Shvydko, et al.2013, Moretti Sala et al., 2018, Sokaras, et al.2013, Kleymenov, Evgeny, et al.2011). Another application of X-ray spectrometer is in lab-based XAS and XES instruments(Jahrman, Evan P., et al.2019, Seidler, G. T., et al.2014). There are different types of X-ray spectrometer based on different types of crystal analyzers such as spherically bent crystal analyzer (SBCA)、cylindrical bent crystal analyzer (CBCA) and so on. We focus here on the case of Rowland circle geometries with a SBCA.

However, in order to provide high-quality experimental conditions for users, it often takes a lot of time and manpower to adjust and optimize the X-ray spectrometer due to its complex mechanical structure and many degrees of freedom. In the face of changing light spots via X-ray area detectors or pulses/photon counts via counting detectors, the staff needs to rely on rich experience in beam modulation to adjust the multidimensional motor for detector and crystal analyers in its travel range one by one, until the sample in the experimental station can obtain the required light flux, light spot shape and light spot size. The entire beam tuning process usually takes several hours, or even longer, resulting in delays to the user's valuable laboratory time. Especially for the experimental line station of X-ray spectrometer based on synchrotron radiation. The reboot after shutdown every year, the machine research every weeks, and the GAP adjustment of line station inserts will all change the electron orbit in the storage ring, resulting in the change of line station light source, thus affecting the quality of the experiment. In many cases, this change can be followed irregularly, which makes it difficult to quickly restore the X-ray spectrometer to the optimal state. How to quickly optimize/re-optimize the X-ray spectrometer to the best state is a serious challenge facing the experimental station of high resolution X-ray spectroscopy.

Due to its high complexity and many degrees of freedom, X-ray spectrometer is difficult to be modeled by analytical mathematical formula. Therefore, rapid optimization of X-ray spectrometer is also a difficult problem faced by major X-ray spectrometer experimental stations. In recent years, with the introduction of automatic optimization technology in synchrotron radiation and other large scientific equipment automatic beam modulation field. For example, in 2015, Dr. Du Yonghua proposed to use genetic algorithm combined with OMEA method to achieve automatic global optimization of synchrotron radiation beam lines (Xi, et al.2015). In 2017, he proposed to replace genetic algorithm with differential evolution algorithm (Xi, et al.2017), which was successfully applied to the beam modulation system of XAFCA line station of Singapore light source. The method developed by Dr. Du has also been applied in the field of X-ray imaging (Chen, et al.2017). In 2020, NSLS-II light source will also introduce intelligent technology (Campbell, et al.2020). Optimization algorithm has also been applied in beam line design (Korchuganov, et al.2018, Taheri, et al.2019). The development of the above fields has a good reference for the adjustment of X-ray spectrometer. In this paper, an automatic adjusting program of X-ray spectrometer based on global optimization algorithm is developed.

# 1. Introduction of global optimization algorithm and its selection for X-ray spectrometer automatic adjustment

Optimization algorithm is widely used in various fields of scientific research and production. The rational application and improvement of optimization algorithm can even play a great role in promoting the development of disciplines. For example, Calypso (Wang et al., 2012) adopted the particle swarm global optimization algorithm, and Uspex (Glass et al., 2006) adopted the genetic global optimization algorithm, both of which have achieved success. In practical application, the analytic description, analytic gradient and analytic Hessen matrix can not be provided due to a considerable part of scientific research and production problems. In the field of optimization algorithms, these optimization problems that cannot provide analytic gradient are usually called black box optimization problems. The word black box refers to the fact that the information of analytic gradient is not available, but not that the problem algorithm itself is black box.

The adjustment of X-ray spectrometer is a typical black box optimization problem. In this paper, the convex optimization and nonlinear optimization, the dependence of gradient information, global optimization and local optimization are analyzed, and it is pointed out that the global (nonlinear) optimization algorithm which does not depend on gradient information should be used in the X-ray spectrum adjustment.

On the first hand, optimization problems can be divided into convex optimization and nonlinear optimization, in which the optimization of convex functions on convex sets is convex optimization, and the rest are nonlinear optimization problems. A prominent feature of convex function optimization is that the local minimum point is even the global minimum point, so convex function has developed its unique series of efficient optimization algorithm, which is called convex optimization. Black box optimization is not necessarily nonlinear optimization, can be defined everywhere, even can be convex optimization. There is no doubt that convex optimization algorithm is more efficient for convex optimization problems, but when it is impossible to accurately judge whether the black box problem is convex, more general nonlinear optimization algorithm can be a choice. In the second aspect, the optimization algorithm can be divided into the gradient information dependent optimization and the gradient information independent optimization. If the amount of calculation for the black box problem is quite small, the numerical gradient can be calculated by different difference schemes, so the gradient-dependent optimization method can be applied. When it comes to the adjustment of X-ray spectrometer, the time of each step is mainly the motor movement time and the detector acquisition time, which is not huge. However, it is difficult to calculate the numerical gradient in the process of X-ray spectrometer adjustment. The third aspect of nonlinear optimization algorithm can be divided into local optimization algorithm and global optimization algorithm, convex optimization because of its characteristics, local optimization is global optimization. If only minor correction of the spectrometer attitude is made, the local optimization algorithm is sufficient. However, if the initial spectrometer attitude is not good enough and the objective function (such as the negative value of SDD detector count) is complex, then the global optimization algorithm is more suitable. It happens that most of the global optimization algorithms are independent of gradient information, and the second and third aspects need to reach agreement. Therefore, it is a reasonable choice to choose the global (nonlinear) optimization algorithm independent of gradient information for the adjustment of X-ray spectrometer.

In the following paper, different optimization algorithms used by shell have different optimization results, and the same optimization algorithm has been run several times with certain deviation, which may cause some misunderstanding. The convergence problem of global optimization algorithm related to it is described here. First of all, the concept of Global Convergent in the local optimization algorithm refers to that the same minimum point can be searched stably for all given different initial structures, rather than the global minimum point can be searched for all different initial structures. The convergence of some global optimization algorithms we used is analyzed below. There are two kinds of global optimization algorithms, one is deterministic global optimization algorithm, the other is random and heuristic global optimization algorithm. Theoretically proving the convergence of the algorithm. The deterministic optimization algorithm usually has a theoretical proof under certain conditions. For example, adaptive grid orientation search (MADS), one of the excellent nonlinear black box optimization algorithms, is an improved form of generalized pattern search (GPS). Adaptive grid orientation search is an abstract algorithm with different implementation modes. The Nomad software defaults to the MADS instantiation of Ortho-Mads (Audet et al., 2006). Due to its development history from classical simplicial local optimization algorithms, MADS has been classified as a local optimization algorithm such as Neumaier's web page, Rois's paper (Rios & Sahinidis, 2013). The improvement of the poll size and the poll direction makes NOMAD have a stronger ability to leap out of the local optimal solution (global optimizer) compared with simplex and generalized pattern search (Audet & Dennis Jr., 2006). Therefore, it is also classified as a global optimization algorithm (Nieuwoudt & Massoud, 2005). Audet et al. presented different levels of convergence of abstract MADS (Audet & Dennis Jr., 2006), and proposed that the lower triangular implementation of MADS, LTMADS, could produce an asymptotically compact set with probability 1 for screening directions. Another example is Direct algorithm (Finkel & Kelley, 2006). Finkel et al. proved that under certain conditions, the optimal solution sequence generated by the algorithm converges to the KKT point (that is, satisfies the KKT condition). Compared with deterministic global optimization algorithms, many random and heuristic global optimization algorithms only guarantee convergence probabilistically or only have heuristic, and there is no complete research on convergence. In the review of Pardalos et al. (Pardalos et al., 2000), many "Stochastic approaches" can only provide "probabilistic convergence guarantee", and tabu search and genetic algorithm "remain heuristics" for global optimization problems. Taking ISRES (Improved Stochastic Ranking Evolution Strategy, Constrained Evolutionary Optimization) algorithm (Runarsson & Yao, 2000) as an example, ISRES is a constrained evolutionary algorithm and is classified as a global non-derivative optimization algorithm by NLOPT nonlinear optimization library. In the algorithm, a mechanism is designed to escape from the local optimal solution, but the global convergence of the algorithm, even under certain conditions, has not been proved strictly.

Even if the optimization algorithm ensures the convergence under certain conditions, it is often difficult to verify the theoretical conditions in practical application. The test set of the algorithm can reflect the performance of the algorithm in practical application. The tests of the global non-derivative optimization algorithm in the NLOPT library can be found in the literature (Kumar et al., 2016). Of course, the literature also contains the tests of the global derivative optimization algorithm and the local optimization algorithm in the NLOPT. Can see optimization algorithm is not effective in all the problems, and for the same optimization algorithm is run on the same issue, because of the initial value, the algorithm introduced in different random seed (radom seeds) and the algorithm for reasons of his own, and the optimization results will have certain distribution range, strict tend not to get consistent results. The size of the distribution varies from algorithm to algorithm. Horizontal comparison of a class of constrained evolutionary optimization algorithms is presented in the literature (Huang Tianyun, 2008). It can be seen that although the algorithm can basically converge to the same result in the simple problem. However, for G2 problem with many variables and complex topological structure of the objective function, the algorithm (including ISRES) cannot achieve stable convergence to the same result every time.

Although non-derivative global optimization algorithms are still under development, there are still some problems in terms of convergence. However, in practice, we find that the non-derivative global optimization algorithm still plays a vital role in the adjustment process of X-ray spectrometer. According to the description above, we pretend to adopt the deterministic global optimization algorithm in X-ray spectrometer automatic adjustment.

## 2. Development and online test of Automatic adjustment process

As mentioned above, X-ray spectrometer is mainly used for high-energy resolution X-ray spectroscopy experiments based on synchrotron radiation and for XAS or XES experiments based on laboratory X-ray sources. The automatic adjustment system in this paper is developed on laboratory XAS equipment, and can also be transplanted to spectrometer on synchrotron radiation beamlines. The hardware equipment in this paper (Fig.1) includes three components: Mo target X-ray source, analytical crystal and detector. The source of X-ray is fixed, and the adjustment dimension mainly includes the three-dimensional translation and one-dimensional rotation of the analysis crystal and detector, totally 8 dimensions. The analytical crystal is Si[5,3,1] plane with a radius of curvature of 500mm. The detector we used is Ketek silicon drift detector (SDD) with a photon sensitive area of 80mm$^2$. Translation and rotation are realized by the stepper motor, and every two motors are controlled by a independent motion controller. All stepper motors can move at the same time. For the convenience of description, we will use XYZ for translational motion, R for rotation, C and D for Crystal and Detector respectively.

We adopt the global optimization algorithm to develop the automatic regulation system. The objective function F is the negative value of the pulse count of the SDD detector. The minimum value obtained by optimization is the maximum value of the detector pulse count, and the positions of each motor are the optimized attitude of the spectrometer to make the light source, the analytical crystal and the detector located on the Rowland circle. To prevent the detector from getting too close to the crystal to pursue an unnormal high pulse count, we attached collimating tubes to it. The objective function can have other options such as the spot shape parameters obtained for the area detector and the photon count within the spot, as well as regulation of the Rowland circle geometry provided with the introduction of computer vision. Using multi-objective optimization algorithms to consider multiple objective functions at the same time is one of the important aspects of subsequent upgrades. This paper is limited to one-dimensional detector counting objective function currently.

In the first step, we obtain the corresponding detector count through the scanning of two dimensional degrees of freedom. Based on this, two dimensional numerical objective functions are obtained by numerical interpolation. Based on this, we conduct a virtual two dimensional optimization test. Fig.2 shows the numerical functions of $F(X_D, Y_D)$ and $F(X_D, Z_D)$. Using the numerical objective function, we can improve the development efficiency and save the on-line testing time. Second, we performed online test. The process of the motor moving to the specified position to collect the detector data is an optimized step. In order to save time, the stepper motor should move to given position simultaneously. In order to fully test the robustness of the automatic adjustment system, we set the motion range to 80mm and 80 degrees. The system can optimize the attitude of the spectrometer quickly and automatically under such a wide boundary condition, which shows its global optimization robustness. In practice, we found that the optimization of 4-5 dimensional can be completed in dozens to hundreds of rounds. All 8-D degrees of freedom can be adjusted through 2-3 optimizations of the spectrometer. We have done several multidimensional optimizations, the number of steps required for optimization and the maximum count obtained by optimization are shown in Tab.1. The optimization process is shown in Fig.3. It is worth noting that since the tests correspond to different energy positions of absorption spectra of different elements XAS experiments, the final optimized detector counts are not the same. In addition, sometimes the adjustment process is carried out in the case of preventing samples, so the detector count after sample absorption will be significantly reduced, especially for the energy above the absorption edge. After the attitude of the spectrometers was automatically adjusted, the monochromatic light was confirmed through the multi-channel mode of SDD detector, and the spectrum was further compared with the standard one obtaining from synchrotron radiation beamline to confirm the correct attitude of the spectrometer.

The software is written by the C++ language binding of Qt. C++ language through the serial command control the motion controller. The XIA detector driver function is called to control the detector for data acquisition. The Widget software interface prepared is shown in Fig.4. Parameters such as the scope of the dimension to be optimized and the condition for the end of optimization can be set. See Figure 5 for the software flow chart.

The optimization effect of X-ray spectrometer is finally realized to the spectrum collected by the spectrometer. We listed the comparison of laboratory XAS collected by us and standard spectra of different element Fe (Fig. 6), Co (Fig. 7) and Cu(Fig.8) . The Fe XAS data collected by us are in good agreement with standard one in energy space, K space and R space. It is worth noting that because the resolution of the monochromatic light obtained by the laboratory X-ray spectrometer is lower than that obtained by the synchrotron radiation monochromator, the broaden is usually larger than that obtained by the synchrotron radiation. In addition, the smaller Bragg angle of the crystal analyzer used for experiment, the larger energy resolution of the spectrometer is obtained. Therefore, the energy resolution of the spectrometer Tab.2 is calculated via ray tracing simulation. When the XAS data of Cu is collected via Si[5,3,1] crystal analyzer, the energy resolution estimated via ray tracing simulation is as low as 8.7eV, so it cannot be directly compared with the standard spectrum(SR in Fig.8), but it is in good agreement with the broadened standard spectrum(SR10eV in Fig.8).

Tab. 1 Optimization steps and pulse count of SDD detector.

| optimization dimension | Total step number(pulse count of detector) |
|---|---|
| 2 | 34steps(count10.5 thousand) |
| 5 | 80steps(count 9.866 thousand) |
| | 44steps(count19.1thousand) |
| | 53steps(count 1513) |

Tab. 2 Ray tracing simulated energy resolution FWHM of Si[5,3,1] at different bragg angle. Source spot size is 0.4mm*0.4mm.

| Fe 74.38 deg | 1.24eV |
|---|---|
| Co 50.12 deg | 6.76eV |
| Cu 45.92 deg | 8.70eV |

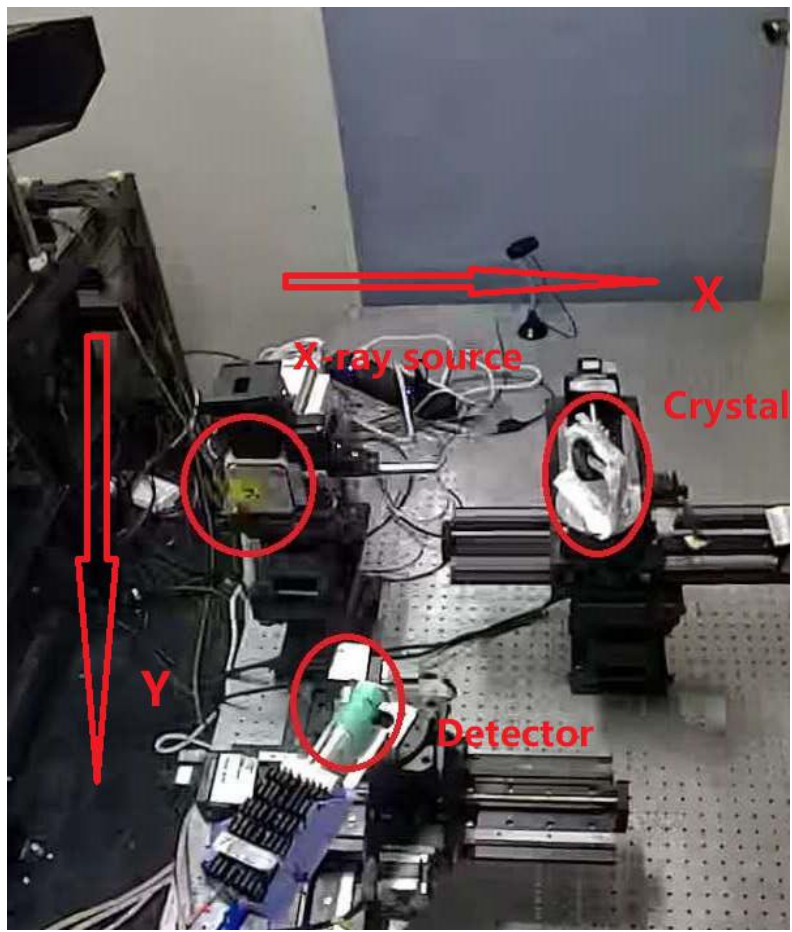

Fig. 1 X-ray absorption experiment equipment based on Mo X-ray tube.

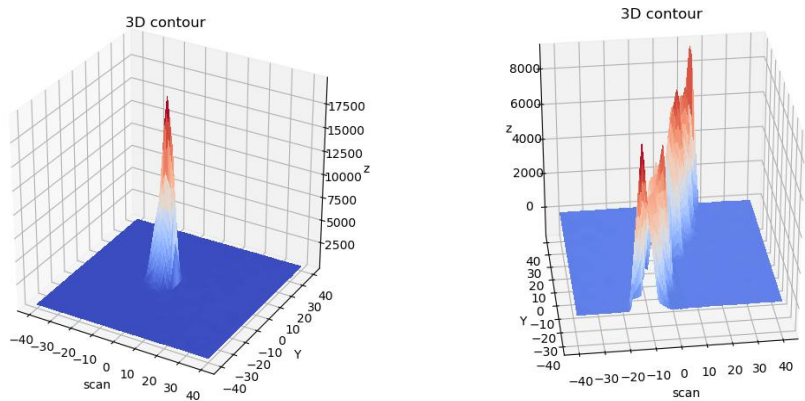

Fig. 2 numerical objective function $F(X_D, Y_D)$ and $F(X_D, Z_D)$ based on scanned experimental data.

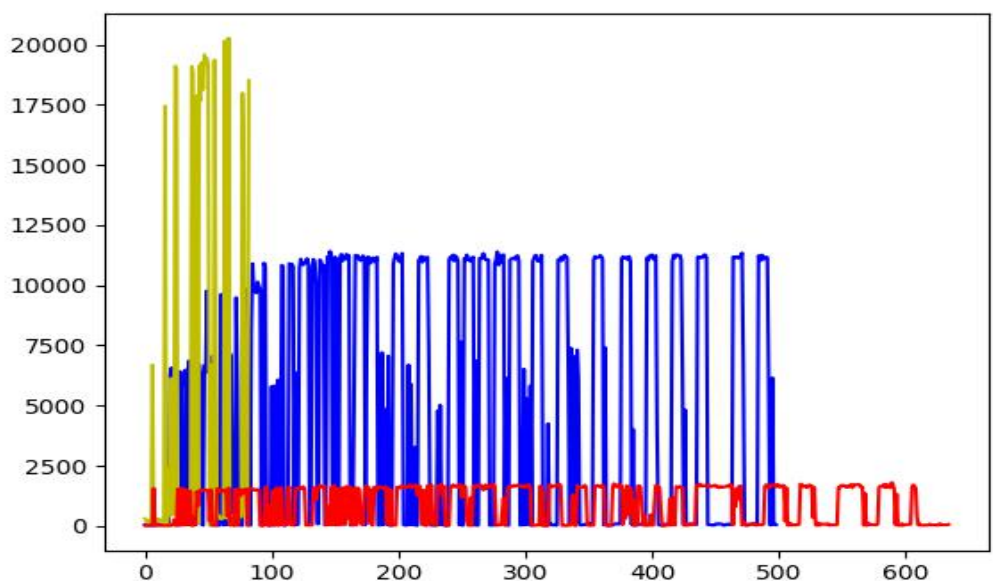

Fig. 3 Several X-ray spectrometer Optimization procedures, the abscissa is the optimization step index, and the axial coordinate is the pulse count of SDD detector.

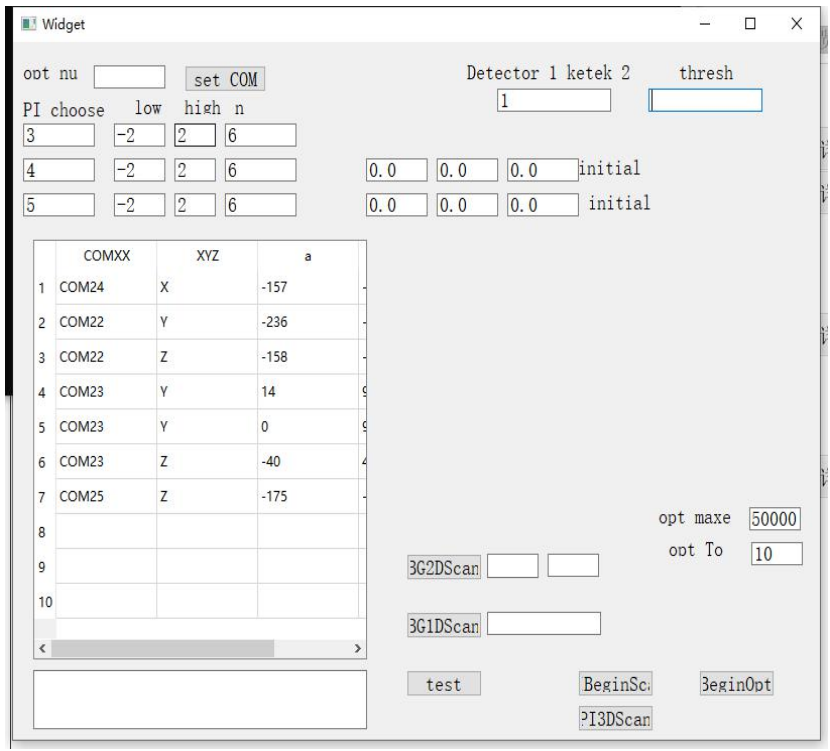

Fig. 4 Qt GUI of X-ray spectrometer automatic adjustment.

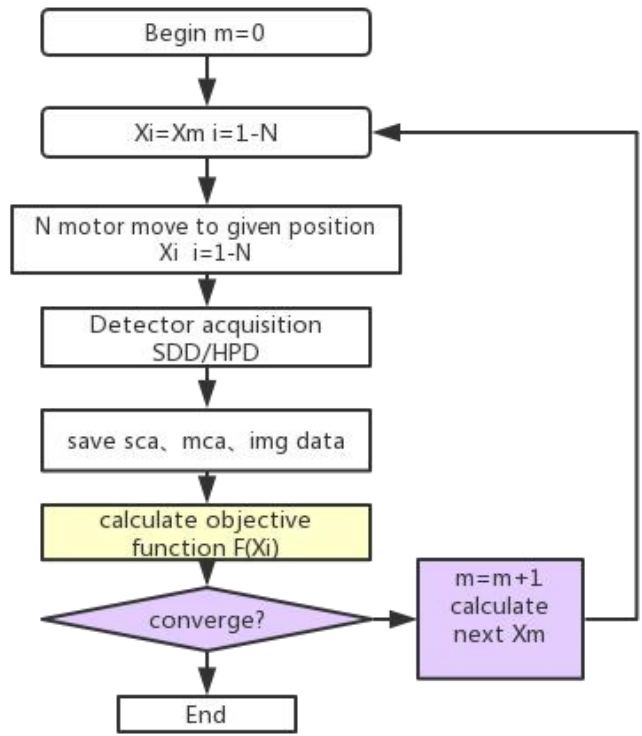

Fig. 5 X-ray spectrometer automatic adjustment software flow chart.

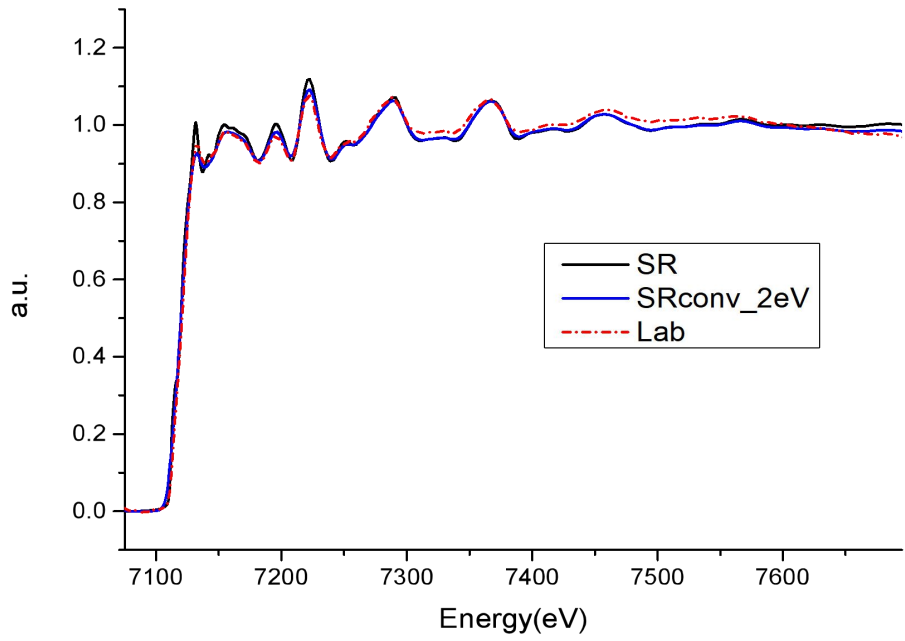

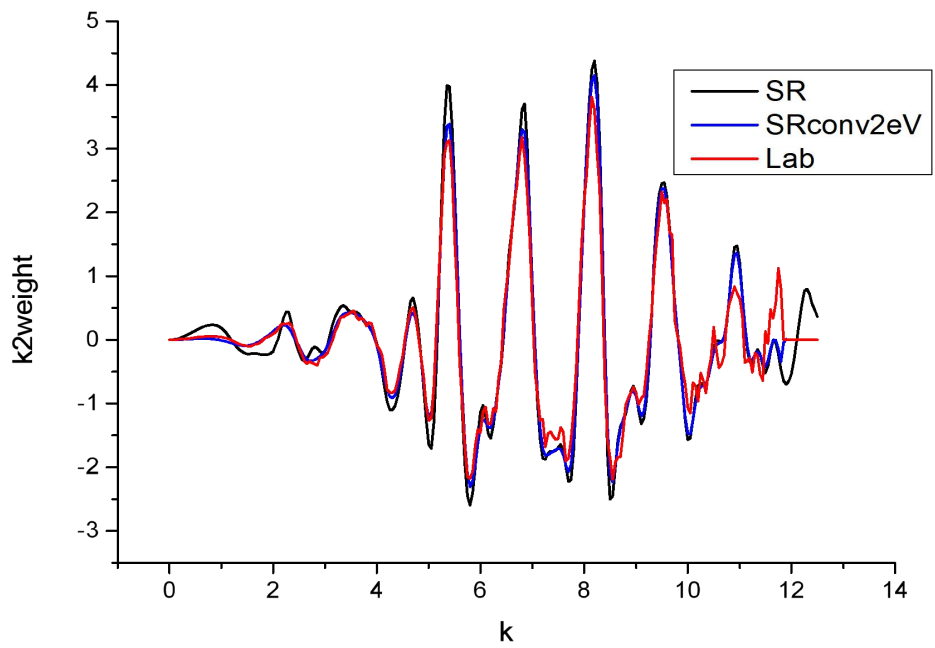

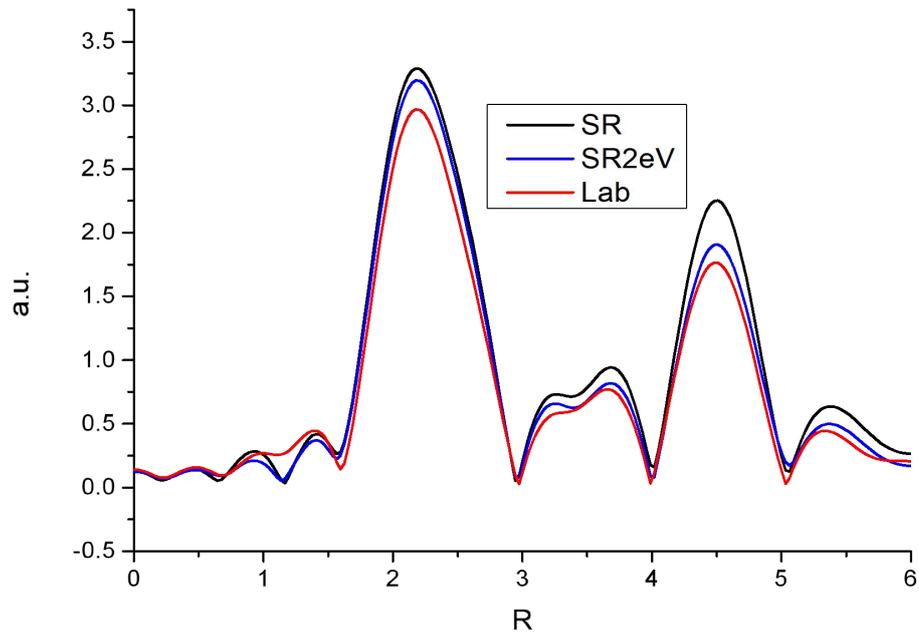

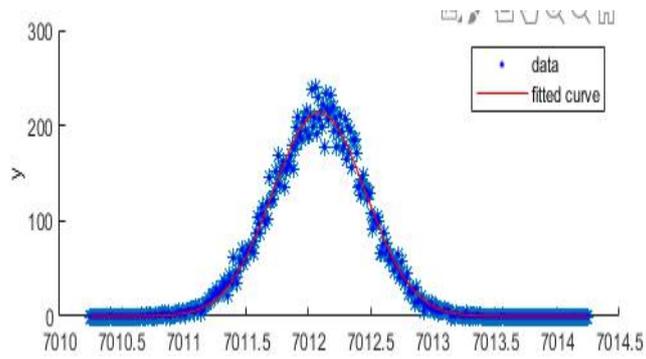

Fig. 6  Comparison of laboratory XAS and synchrotron radiation standard spectra of Fe in energy、K、R Space. Ray tracing simulated energy resolution.

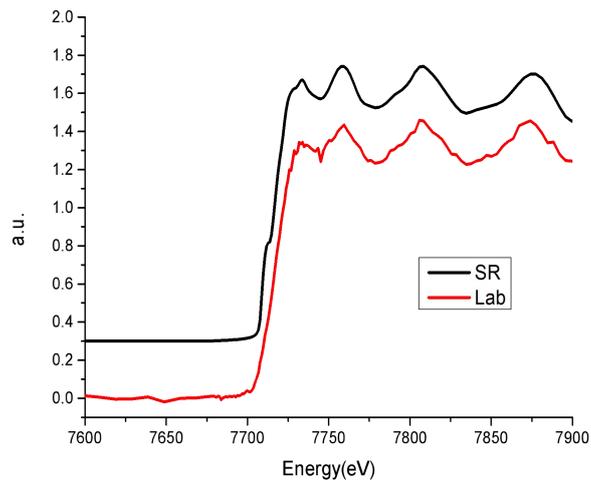

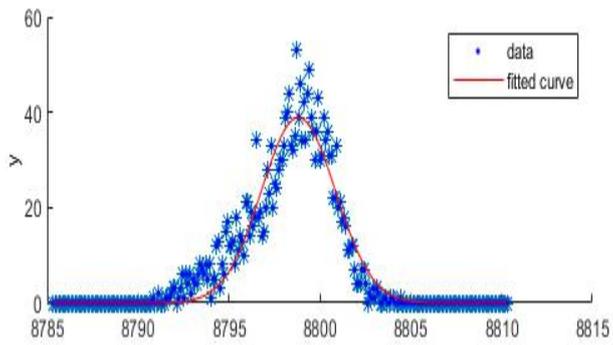

Fig. 7 Comparison of laboratory XAS and synchrotron radiation standard spectra of Co in energy Space. Ray tracing simulated energy resolution.

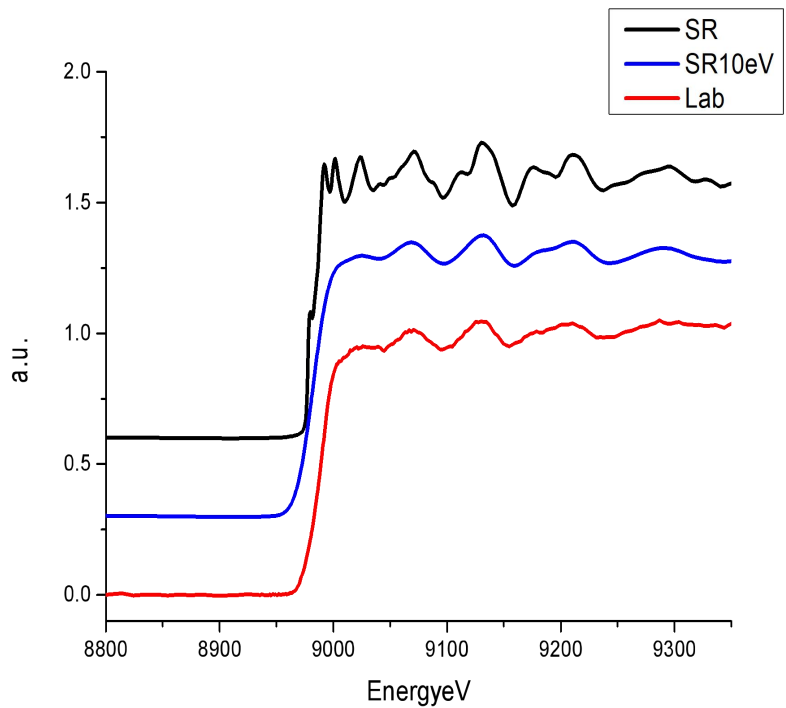

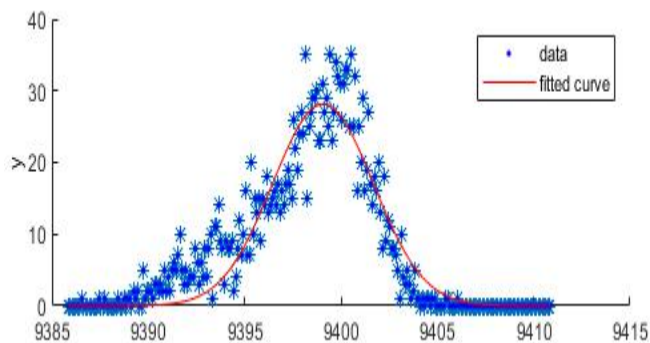

Fig. 8 Comparison of laboratory XAS and synchrotron radiation standard spectra of Cu in energySpace. Ray tracing simulated energy resolution.

## 3.Conclusion and Outlook

In this paper, the global optimization algorithm is applied to realize the automatic optimization and adjustment system of X-ray spectrometer. The system can search automatically within the set range, accurately find and converge to the optimal solution, so as to realize the automatic optimization and adjustment of multi-dimensional freedom of X-ray spectrometer. Optimization process is about dozens to hundreds of steps. The efficiency and robustness of the automatic optimization and regulation system of X-ray spectrometer are related to many factors, such as the performance of optimization algorithm, the response speed of motion controller, the motion speed and positioning accuracy of motor, and the selected optimization index. Improving the performance of each link is the basis to improve the efficiency of the entire system, because the efficiency of the system has a large room for improvement. In addition to the pulse count of SDD detector, it is of great significance to take the spot shape and geometric position of the spectrometer element based on computer vision as the optimization objection to improve the globality and robustness of the system, which is also the future development direction of the system. The work in this paper provides a feasible scheme for the adjustment of X-ray spectrometer, and has important reference value for such experimental systems based on synchrotron radiation source and laboratory X-ray source.